\documentclass[10pt,pra,twocolumn,superscriptaddress,showpacs,showkeys]{revtex4-1}

\usepackage[dvips]{graphicx}
\usepackage{amsmath,graphics,amsfonts,bm,amssymb}
\usepackage{times}

\begin{document}

\title{Tunable interaction of superconducting flux qubits in circuit QED}

\author{Zheng-Yuan Xue}\email{zyxue@scnu.edu.cn}
\affiliation{Guangdong Provincial Key Laboratory of Quantum Engineering and Quantum Materials,  and School of Physics\\ and Telecommunication Engineering, South China Normal University, Guangzhou 510006, China}

\author{Ya-Fei Li}
\affiliation{Guangdong Provincial Key Laboratory of Quantum Engineering and Quantum Materials,  and School of Physics\\ and Telecommunication Engineering, South China Normal University, Guangzhou 510006, China}

\author{Jian Zhou}
\affiliation{Anhui Xinhua University, Hefei, 230088, China}
\affiliation{National Laboratory of Solid State Microstructure, Nanjing University, Nanjing 210093, China}
\affiliation{Guangdong Provincial Key Laboratory of Quantum Engineering and Quantum Materials,  and School of Physics\\ and Telecommunication Engineering, South China Normal University, Guangzhou 510006, China}

\author{Yu-Mei Gao}
\affiliation{Zhongshan College, University of Electronic Science and Technology of China, Zhongshan, 528402, China}

\author{Gang Zhang} \email{zhanggang@wxc.edu.cn}
\affiliation{School of Mechanical and Electronic Engineering, West Anhui University, Lu'an 237012, China}

\date{\today}

\begin{abstract}
We propose to  implement tunable interaction of superconducting  flux qubits with cavity-assisted interaction and strong driving. The qubits have a three-level Lambda configuration, and the decay of the excited state will be greatly suppressed due to the effective large detuning. The implemented interaction is insensitive to the cavity field state and can be controlled by modulating the phase difference of the driving fields of the qubits. In particular, our scheme is based on the typical circuit QED setup and thus will provide a simple method towards the tunable interaction of superconducting qubits. Finally, we consider the generation of two and four qubits entangled  states with the constructed interaction under the influence of typical decoherence effects.
\end{abstract}



\maketitle


Recently, with highly developed fabrication techniques and
newly designed architecture,  superconducting qubit is recognized as  one of the most promising candidates for implementation of the on-chip quantum computation \cite{sq,sqyou,sq1,sq2}. To implement quantum networks and distributed quantum information processing, cavity-assisted coupling is preferred \cite{c1,c2,cqed1,cqed2,cqed3,c3,c4,c5}, where a microwave photon can be an ideal flying qubit \cite{flying}. Moreover, this architecture has several practical advantages including strong coupling strength, immunity to
noises, and suppression of spontaneous emission. As it is well known, the ability of controlling qubit-qubit and qubit-cavity interactions is generic to
all physical implementation of quantum computer and simulators \cite{c6,c7,c8,qubitcavity}. However, the cavity-induced interaction is of the always-on nature, and thus one needs to find practical ways of switching and/or tuning this interaction.

Conventionally, in the circuit QED scenario, the qubit-cavity is effectively tuned off by dispersive interaction \cite{cqed1}. This requires one can tune the energy splitting of the  superconducting qubit and/or the cavity. The more severe problem is that this switch will become less effective in the case of more than single qubit in the cavity. In this case, qubits can be effectively interacted with each other by exchange the virtual photon of the cavity \cite{virtual1,virtual2}, and thus may sweep the switch effect. Therefore, one needs more accurate tunable qubit-cavity interaction. With this tunable qubit-cavity interaction, one may achieve tunable qubit-qubit interaction as the coupling strength  of which is determined by the qubit-cavity coupling strengths \cite{virtual1}.  However, previous investigations \cite{tune1,tune2,tune4,tune5,tune6,tune7,tune8} are based on  increasing the complexity of their setups to achieve this tunability, e.g. introduce a SQUID serves as a tunable coupler  and thus will inevitably introduce additional noise for the qubits.  Moreover, there is no investigation in tunable interaction of superconducting qubits in a cavity.

Here, we propose to directly implement tunable interaction for superconducting qubits from the always-on qubit-cavity interaction in circuit QED. To obtain interaction among qubits, we use the dispersive qubit-cavity interaction together with microwave-fields-induced strong driving.  In particular, the obtained coupling strength of the superconducting qubits can be tuned in a continuous way by modulating the phase difference of the effective qubit-cavity interaction.  The phases come from the external driving fields which can be tuned individually and thus will provide a simple method towards the tunable interaction of superconducting qubits. The superconducting qubits considered here are flux qubits with three energy levels in effect \cite{3level1,3level2,yxl}.  The two driving fields of a qubit compete with the qubit-cavity interaction and lead the whole interaction to a collective form, i.e. collective qubit-states-dependent displacement of the cavity field. In this way, the cavity state traverses cyclically and returns to its original state at some intervals and leaves an geometric phase on the collective states of the involved qubits \cite{solano,e1,zhuwang,xue1,xue2,xue3,zzm}, which avoids the cooling of the cavity to its exact ground state beforehand and results in high-fidelity operation on the qubits.

\begin{figure*}[tb]
\begin{center}
\includegraphics[width=1.7\columnwidth]{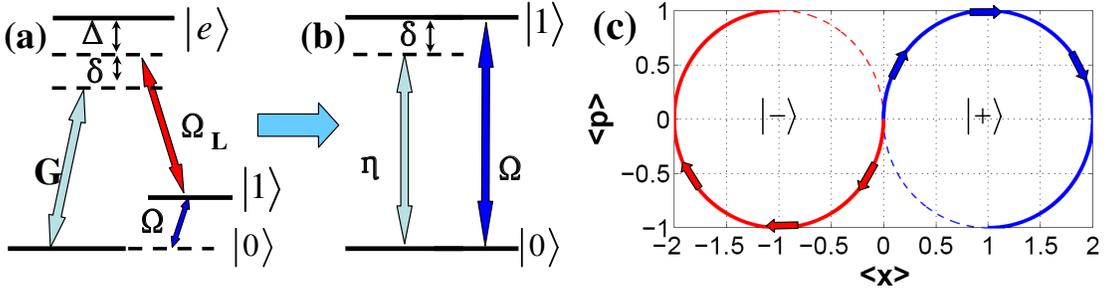}
\end{center}
\caption{Coupling configuration of our proposal.  (a) The Delta type of transition are coupled by the cavity field and two microwave fields. (b) The effective model of (a) is the circuit QED coupling with driving. (c) The closed trajectory results from the spin-dependent force with the unit being $\sqrt{2}\eta/\delta$.}
\label{f1}
\end{figure*}


We now turn to our proposal. The superconducting qubits are flux qubits with three energy levels in a Lambda configuration labelled by $|0\rangle$,  $|1\rangle$ and  $|e\rangle$, as illustrated in Fig. \ref{f1}(a). When the external magnetic flux $\Phi_e$ pierced through the qubit loop is deviated from the symmetry point with $f=\Phi_e/\Phi_0=1/2$, the selection rules of the qubit will be broken \cite{3level1}, allowing for the coexistence of one- and two-photon processes, i.e. microwave-induced transitions between any two energy levels are possible. In our implementation, $\left\vert 0\right\rangle $ and $\left\vert 1\right\rangle $ are encoded as our qubit states, which are coupled with a resonant microwave field with strength $\Omega$. Here, we choose $f$ to be slightly deviated from $1/2$, which leads the qubit energy splitting  $\omega_q$ to be in the sub-GHz level  \cite{3level1} and the three levels form  a Delta type of transition configuration.   In addition, the qubits are capacitively coupled to a one-dimensional superconducting cavity forming the typical circuit QED architecture. The transition $\left\vert 0\right\rangle \rightarrow \left\vert e\right\rangle$ ($\left\vert 1\right\rangle \rightarrow\left\vert e\right\rangle $) with frequency $\omega_{e0}$ ($\omega_{e1}$) is coupled by a superconducting cavity (a microwave field) with frequency $\omega _{c}$ ($\omega _{L}$) and coupling strength $G$ ($\Omega_L$). Then, in units of $\hbar =1$, the system is described by
\begin{eqnarray}
H_{S}&=&H_0
+\sum _{j=1}^{N}\left[ {\Omega_j \over 2}
e^{-i \omega_q }|1\rangle\langle 0| +G _{j} a|e\rangle\langle 0| \right.\notag\\
&& \left. + \Omega_{L,j}e^{-i(\omega_L+\phi_j)} |e\rangle\langle 1| + \text{H.c.}\right],
\end{eqnarray}
where $H_0=\omega_c a^{+} a +  \sum_i \omega_i |i\rangle\langle i|$ is the free Hamiltonian where $a^{+}(a)$ is the creation (annihilation) operator of the cavity field and $\omega_i$ is the frequency of $i$th energy level with $i\in\{0, 1, e\}$.
Note that both the $\left\vert 0\right\rangle \rightarrow \left\vert e\right\rangle$ and $\left\vert 1\right\rangle \rightarrow\left\vert e\right\rangle $ couplings are far-off resonant from their transition frequencies  so  that the excited state $|e\rangle$ state can be adiabatically eliminated \cite{feng,e3,e4} and thus suppress the influence of the spontaneous emission effect to our implementation. Within the rotating-wave approximation, the interaction Hamiltonian in the interaction picture,  with respect to $H_0$, can be written  as
\begin{equation}\label{hint}
H_{I}=\sum _{j=1}^{N} {\Omega_j \over 2} \sigma_j^x
+ 2\eta _{j}\left(a\sigma _{j}^{+}e^{i (\delta_{j} t +\phi_j)}+\text{H.c}.\right),
\end{equation}
where   $\sigma _{j}^{+}=\left\vert 1_{j}\right\rangle \left\langle
0_{j}\right\vert $, $\sigma _{j}^{-}=\left\vert 0_{j}\right\rangle
\left\langle 1_{j}\right\vert $, and
\begin{equation}
\eta _{j}={G_{j}\Omega _{L,j} \over 2} \left(\frac{1}{\Delta _{j}+\delta _{j}}+\frac{1}{\Delta _{j}}\right) \simeq  {G_{j}\Omega _{L,j} \over \Delta_j}
\end{equation}
with $\Delta _{j}=\omega _{e1,j}-\omega
_{L,j}$, $\delta _{j}=\omega _{01,j}+\omega_{L,j}-\omega _{c}$  and $\Delta_j \gg \delta_j$. Therefore, we realize the circuit QED system with microwave driving, as shown in Fig. \ref{f1}(b), where the effective cavity-induced coupling $\eta_j$ can be tuned by the $\Omega _{L,j}$  and detunings. In addition, the phase factors $\phi_j$ can be tuned individually.

We next detail our further treatments.
Assuming   $\eta_{j}=\eta$ and $\delta_{j}=\delta$, in the interaction picture with respect to the first term, the interacting Hamiltonian in Eq. (\ref{hint}) reads \cite{solano}
\begin{eqnarray} \label{h1}
&& H_{1}= \eta \sum_{j=1}^N \left[a e^{i(\delta  t+\phi_j)} + a^\dagger e^{-i(\delta  t+\phi_j)} \right] \sigma_{x}^j\\
&+& \eta \sum_{j=1}^N \left[a e^{i(\delta  t+\phi_j)} \left(e^{i\Omega t}|+\rangle_j\langle-|
-e^{-i\Omega t}|-\rangle_j\langle+|\right)+\text{H.c.}\right].\notag
\end{eqnarray}
where $|\pm\rangle=(|0\rangle\pm|1\rangle)/\sqrt{2}$. In the case of strong driving, $\Omega\gg\{\delta, \eta \}$, we can omit the fast oscillation terms with frequencies $\Omega\pm\delta$, i.e. the second line of Hamiltonian (\ref{h1}), and then we obtain
\begin{eqnarray}\label{h2}
 H_{2}=  \eta  \sum_{j=1}^N \left[a e^{i(\delta  t+\phi_j)}
 + a^{\dag} e^{-i(\delta  t+\phi_j)} \right] \sigma_j^x,
\end{eqnarray}
which is in the form of spin-dependent dipole force for the cavity field. This force will lead the cavity state return to its original state periodically in phase space. To see this, we consider the case that only the first qubit is subjected to the force in the $xp$ phase space with $\phi_1=0$. In this case, the Hamiltonian (\ref{h2}) reduces to
\begin{eqnarray}\label{h3}
 H_{3}= \sqrt{2} \eta \left[\cos(\delta  t)x-
\sin(\delta  t)p \right] \sigma^x,
\end{eqnarray}
where we have defined the dimensionless position and momentum operators, $x=(a^\dagger +a)/ \sqrt{2}$ and $p=i(a^\dagger -a)/ \sqrt{2}$. In the phase space, the effect of Hamiltonian (\ref{h3})  is to perform the displacements of
\begin{eqnarray}
(x_0, p_0)\rightarrow ((x_0+x(t))\sigma^x, (p_0+p(t))\sigma^x)
\end{eqnarray}
with $x(t)=\sqrt{2}\eta(1-\cos\delta t)/\delta$ and $p(t)=\sqrt{2}\eta\sin(\delta t)/\delta$, which entangles the cavity with the qubit. Depending on the spin states $|\pm\rangle$, the cavity is displaced to form a closed trajectory, see Fig. \ref{f1}(c) for $x_0=p_0=0$. When incorporating many qubit, this spin-dependent force can be used to generate many-body entangled state of the qubits, i.e., introduce nontrivial interaction among qubits. As $x(t)$ and $p(t)$ are both periodical functions, the cavity will return to its original state once $\delta \tau_n=2n\pi$ with $n=1,2,3...$, i.e., the evolution is insensitive to the cavity field state, and thus one may expect that it is more robust, which mitigates the thermal limitation of  the cavity in engineering quantum states. Therefore, at these time intervals, the net effect of Hamiltonian (\ref{h2}) makes the qubit system pick a geometric phase \cite{zhu2003} and leads to fast geometric gate of the qubits.


We next exemplify the nontrivial interaction with $N=2$. When $\delta \tau_n=2n\pi$ and $\phi_j=0$, the induced evolution operator for the Hamiltonian in Eq. (\ref{h2}) is  \cite{e1}
\begin{eqnarray}\label{u}
U(\theta)=  \exp\left(- i \theta  \sigma_1^x \sigma_2^x \right),
\end{eqnarray}
where $\theta=\lambda \tau_n \cos\varphi_1$ with $\lambda=2\eta^2/\delta$ and $\varphi_1=\phi_1-\phi_2$.  Therefore, the corresponding effective Hamiltonian of the operator in Eq. (\ref{u}) is
\begin{eqnarray}\label{eff1}
 H_{4}= \lambda \cos \varphi_1 \sigma_1^x\sigma_2^x,
\end{eqnarray}
where $\varphi_1$  can be used to tune the effective coupling strength. However, the effective Hamiltonian only holds for the time intervals of $\tau_n$. When $\delta \gg \eta$, the effective Hamiltonian will hold at approximately any time due to the fact that $\tau_n$ will be much smaller than the time scale of the periodicity of the effective Hamiltonian, which is on the order of $1/\lambda$.
Therefore, our scheme is faster comparing with previous implementations involving large detuned qubit-cavity interaction.

For an initial state of $|00\rangle$, governed by the Hamiltonian  in Eq. (\ref{h2}), the final state will be a maximally entangled Bell-type state when $\theta=\pi/4$ as \begin{eqnarray}
|\psi_f\rangle = U\left({\pi\over 4}\right)|00\rangle
={1 \over \sqrt{2}} (|00\rangle+i|11\rangle).
\end{eqnarray}
When $\Omega_{L}=2\pi \times 0.1$ GHz and $\Delta=10\Omega_{L}$, $\eta \simeq G /10=2\pi \times 10$ MHz for a moderate cavity-assisted coupling strength of $G=2\pi \times 0.1$ GHz \cite{cqed3}. When $\Omega\sim\Delta$, the assumption $\{\Omega, \Delta \} \gg \delta$ is well satisfied. In order to derive Eq. (\ref{h2}), we have  discarded the terms in the second line of  Eq. (\ref{h1}), which induce Stark shifts on the states of $|\pm\rangle_j$. In the above example of generating Bell state, this neglection will introduce an infidelity approximately proportional to $(\eta/\Omega)^2$, which is negligible small for large ratio of $\Omega/\eta$. In addition, $\theta=\pi/4$ can be realized by choosing $\delta = 4 \sqrt{n} \eta$, which further leads to $\tau_n=\sqrt{n}\pi/(2\eta)$. Therefore, for $n=1$ ($\delta = 4 \eta$), the minimal generation time of the entangled state is $\tau_1 =\pi/(2\eta)$, which is much faster comparing with the usual large detuning schemes ($\delta \gg \eta$).

\begin{figure}[tbp]
\includegraphics[width=.95\columnwidth]{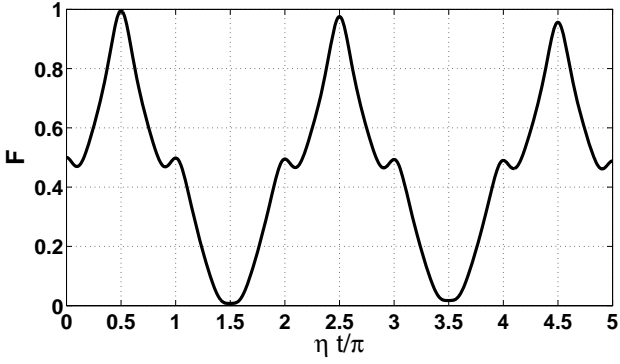} \label{fig2}
\caption{The fidelity dynamics of entangling two  qubits as a function of $\eta t/\pi$  with $\varphi_1=0$, $n=1$ and $\kappa=\gamma_1=\gamma_2=\eta/1000$.}
\end{figure}

The performance of the entangling gate in Eq. (\ref{u}) can be evaluated by considering the influence of dissipation using the  quantum master equation of
\begin{eqnarray}  \label{me}
\dot\rho = i[\rho, H_2]
+\frac \kappa  2  \mathcal{L}(a)  + \sum_{j=1}^2 \left[\frac{\gamma_1}{2} \mathcal{L}(\sigma^-_j) + \frac{\gamma_2}{2}  \mathcal{L}(\sigma^\text{z}_j\right],
\end{eqnarray}
where $\rho$  is the density matrix of the considered system, $\mathcal{L}(A)=2A\rho A^\dagger-A^\dagger A \rho -\rho A^\dagger A$ is the Lindblad operator, $\kappa$, $\gamma_1$  and $\gamma_2$ are the decay rate of the cavity, the decay and dephasing rates of the qubits, respectively. Nowadays, these rates can all be on the order of kHz level \cite{cqed3}, we choose $\kappa=\gamma_1=\gamma_2=\eta/1000=2\pi \times 10$ kHz for demonstration purpose. We evaluate this generation by the fidelity defined by $F=\langle\psi_f|\rho_a|\psi_f\rangle$, with $\rho_a$ being the reduced  density
matrix of the qubits from $\rho$. We plot  $F$ with $n=1$ as a function of dimensionless time $\eta t/ \pi$ with $\varphi_1=0$, where we have obtained a high fidelity of 99.6 \% for $\delta=4\eta$ at $t=\tau_1$ as shown  in Fig. 2.

\begin{figure}[tbp]
\includegraphics[width=.95\columnwidth]{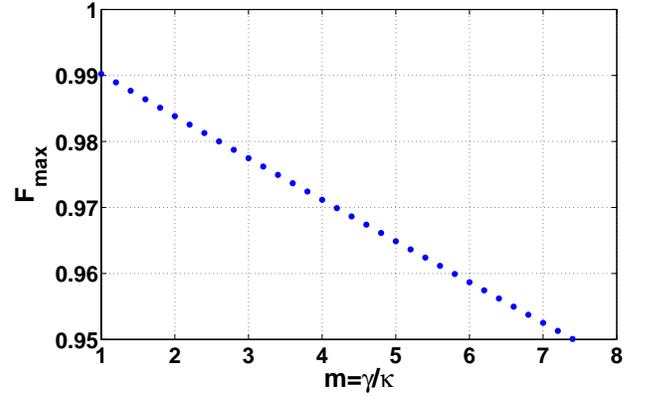} \label{fig3}
\caption{The maximum fidelity of entangling four  qubits for different $m=\gamma/\kappa$ with $\phi_{j,k}=0$, $\gamma_1=\gamma_2=\gamma$, $n=1$ and $\kappa=\eta/1000$.}
\end{figure}


We proceed to consider the scalability  of the effective Hamiltonian  in Eq. (\ref{eff1}).  For the multi-qubit case, the tunability of the corresponding effective Hamiltonian becomes much more complicated as the Hamiltonian will be in the form of
\begin{eqnarray}
H_{5} = \lambda \sum_{1\leq j<k\leq N} \cos\phi_{j,k}\sigma_j^x\sigma_k^x,
\end{eqnarray}
where  $\phi_{j,k}=\phi_j-\phi_k$. In this case, one qubit will interact with all the others and we cannot control all the pairs of qubits in a tunable way in a simple form as in Eq. (\ref{eff1}). Nevertheless, we can use the induced evolution operator to generate multipartite entangled state \cite{ghz,xue}.
That is, for $\phi_{j,k}=0$, when $\delta \tau_n=2n\pi$, the induced evolution operator from Eq. (\ref{h2}) is
\begin{eqnarray}\label{un}
U(\theta)=  \exp\left[- i {\theta \over 2} \left (\sum_{j=1}^N \sigma_j^x\right)^2 \right],
\end{eqnarray}
which can be used to generate GHZ states when $\theta=\pi/4$ \cite{ghz}. For $N$=4, we also evaluate the performance of this generation by numerically integrate  the  quantum master equation, as shown in Fig. 3, where we obtain that the fidelity can be $F_{max}=99 \%$ with  $n=1$, $\gamma_1=\gamma_2=\gamma$ and $\gamma=\kappa=\eta/1000$.  It also shows that with the increase of the qubit decay rate $\gamma=m\kappa$ the maximum fidelity will decrease linearly. Finally, we also note that, if we consider the scenario that each qubit locates in a cavity and different cavities are coupled to form a coupled chain of cavities \cite{cc}, the induced effective Hamiltonian will be
\begin{eqnarray}
H_{6} = \lambda' \sum_{1\leq j \leq N-1} \cos\phi_{j,j+1}\sigma_j^x\sigma_{j+1}^x,
\end{eqnarray}
which can be universal for quantum computation purpose.

In summary, we have proposed to  implement tunable interaction of superconducting  flux qubits with cavity-assisted interaction and strong driving. The decay of the excited state of the flux qubits can be greatly suppressed due to the effective large detuning. The implemented interaction is insensitive to the cavity field state and can be controlled by modulating the phase difference of the driving fields of the qubits. In particular, our scheme is based on the typical circuit QED setup and thus will provide a simple method towards the tunable interaction of superconducting qubits. Finally, we consider the generation of entangled  states from the tunable interaction under typical decoherence effects and show that high-fidelity entangled state can be obtained.

\bigskip

\noindent {\bf Acknowledgments}\\
This work was supported by the NFRPC (No. 2013CB921804), the PCSIRT  (No. IRT1243), the NSF of Anhui Province (No. 1408085MA20), the STPF of Zhongshan (Grnat No. 20123A326), the key project from Educational department of Anhui Province (KJ2015A299), and the fund from NJU (No. M28015).

\end{document}